\documentclass[aps,prl,reprint,groupedaddress,superscriptaddress,preprintnumbers]{revtex4-2}

\usepackage{subfigure}
\usepackage{graphicx,natbib}
\usepackage{bm}
\usepackage{color}
\usepackage{amssymb}
\usepackage[margin=0.6in]{geometry}
\usepackage{multirow}
\usepackage{mathrsfs}
\usepackage{hyperref}

\newcommand{\BB}{\bm{B}}

\newcommand{\UU}{\bm{U}}

\newcommand{\SSSS}{\mbox{\boldmath ${\sf S}$} {}}

\newcommand{\meanrho}{\langle\rho\rangle}
\def\cs{c_{\rm s}}
\newcommand{\nab}{\bm{\nabla}}
\def\Rm{{\rm Re}_{_\mathrm{M}}}
\newcommand{\bra}[1]{\langle #1\rangle}
\def\cs{c_{\rm s}}

\def\kB{k_{\rm B}}

\newcommand{\meanBB}{\langle\mbox{\boldmath $B$}\rangle{}}{}
\newcommand{\alphaem}{\ensuremath{\alpha_{\rm em}}}

\newcommand{\meanmufive}{\bra{\mu_5}}
\newcommand{\meanEMF}{\bra{\mbox{\boldmath{${\cal{E}}$}}}{}}{}

\newcommand{\mnras}{Mon.\ Not.\ Roy.\ Astron.\ Soc.}

\begin{document}

\title{Chiral anomaly and dynamos from 
inhomogeneous chemical potential fluctuations}

\preprint{NORDITA-2023-041}

\author{Jennifer~Schober}
\email{jennifer.schober@epfl.ch}
\affiliation{Institute of Physics, Laboratory of Astrophysics, \'Ecole Polytechnique F\'ed\'erale de Lausanne (EPFL), 1290 Sauverny, Switzerland}

\author{Igor~Rogachevskii}
\affiliation{Department of Mechanical Engineering, Ben-Gurion University of the Negev, P.O. Box 653, Beer-Sheva 84105, Israel}
\affiliation{Nordita, KTH Royal Institute of Technology and Stockholm University, 10691 Stockholm, Sweden}

\author{Axel~Brandenburg}
\affiliation{Nordita, KTH Royal Institute of Technology and Stockholm University, 10691 Stockholm, Sweden}
\affiliation{The Oskar Klein Centre, Department of Astronomy, Stockholm University, AlbaNova, SE-10691 Stockholm, Sweden}
\affiliation{School of Natural Sciences and Medicine, Ilia State University, 0194 Tbilisi, Georgia}
\affiliation{McWilliams Center for Cosmology and Department of Physics, Carnegie Mellon University, Pittsburgh, Pennsylvania 15213, USA}

\date{\today}

\begin{abstract}
In the standard model of particle physics, the
chiral anomaly
can occur in relativistic plasmas
and plays a role in the early Universe, proto-neutron stars, heavy-ion collisions,
and quantum materials.
It gives rise to a magnetic instability if
the number densities of left- and
right-handed electrically charged fermions
are unequal.
Using direct numerical simulations, we show this can result
just from spatial fluctuations of the chemical potential,
causing a chiral dynamo instability,
magnetically driven turbulence, and ultimately
a large-scale magnetic field through the magnetic $\alpha$ effect.
\end{abstract}

\maketitle


The standard model of particles predicts the occurrence
of a macroscopic quantum phenomenon named the 
``chiral magnetic effect'' (CME) \citep{Vilenkin:80a} in plasmas 
with high-energy fermions.
The CME has been derived using different approaches
\citep{RW85,Tsokos:85,Frohlich:2000en,Frohlich:2002fg,Fukushima:08,Son:2009tf,AlekseevEtAl1998}.
It implies an electric current along a magnetic
field, which arises if there is an asymmetry in the chemical
potentials of left- and right-handed fermions, $\mu_\mathrm{L}$ and 
$\mu_\mathrm{R}$, respectively, i.e., if
the chiral chemical potential does not vanish, 
$\mu_5 \equiv \mu_\mathrm{L} - \mu_\mathrm{R}\neq 0$.
Using chiral magnetohydrodynamics 
\citep{AlekseevEtAl1998,GI13,RogachevskiiEtAl2017,HattoriEtAl2019}, it has been shown that the CME
leads to chiral dynamo instabilities
\citep{BFR15,GKR15,KH14,K16,SchoberEtAl2017,SBR19,SBR20,JS97,BFR12}, which can amplify the magnetic energy by many orders of magnitude.
The CME and the chiral dynamo instabilities have
relevance for the early Universe 
\citep{JS97,BFR12,BRS12,DvornikovSemikoz2017,BSRKBFRK17},
proto-neutron stars
\citep{Dvornikov:2015lea,YA16,Sigl:2015xva}, quark-gluon
plasmas \citep{AY13,HKY15,TW15}, heavy ion collisions \citep{KH14,K16},
and for quasi-particles in new materials such as 
graphene and Dirac semimetals \citep{MS15}.
Recently, it has been shown 
through direct numerical simulations (DNS)
that the chiral dynamo instability 
occurs not only in systems with initially homogeneous $\mu_5$, but can develop 
even from spatial fluctuations of $\mu_5$ with zero mean
\citep{SchoberEtAl2022a,SchoberEtAl2022b}.
Producing chiral asymmetry from an initially vanishing 
$\mu_5$ remains, however, a key question.

In the present study, we demonstrate through DNS
that an initially vanishing chiral asymmetry
can be produced from an inhomogeneous
chemical potential, $\mu=\mu_\mathrm{L} + \mu_\mathrm{R}$,
with just spatial fluctuations ($\nabla \mu \neq 0$).
A conversion between $\mu$ and $\mu_5$ is possible
due to what is known as the chiral separation effect
\citep{KH14,K16} and results in a production of 
large values of $\mu_5$, which are sufficient for the
excitation of a chiral dynamo instability.
This leads to the production of magnetically driven
turbulence and mean-field dynamo action,
which causes the generation of a large-scale magnetic field.


To study these effects, we consider
the following set of equations for an effective 
description of a plasma composed of chiral 
electrically charged fermions with the CME and the chiral separation effect:
\begin{eqnarray}
  \frac{\partial \BB}{\partial t} &=& \nab \times \left[{\UU}\times {\BB}
  + \eta \, \left(\mu_5 {\BB} -\nab   \times   {\BB}
    \right) \right],
\label{ind-DNS_CSE}\\
  \rho{D \UU \over D t}&=& (\nab   \times   {\BB})  \times   \BB
  -\nab  p + \nab  {\bm \cdot} (2\nu \rho \SSSS) ,
\label{UU-DNS_CSE}\\
  \frac{D \rho}{D t} &=& - \rho \, \nab  \cdot \UU ,
\label{rho-DNS_CSE}\\
  \frac{D \mu}{D t}  &=& -\mu \, \nab  \cdot \UU - C_\mu (\BB {\bm \cdot} \nab)  \mu_5 - \mathcal{D}_\mu \, \nabla^4 \mu ,
\label{mu-DNS_CSE} \\
  \frac{D \mu_5}{D t} &=& -\mu_5 \, \nab  \cdot \UU - C_5 ({\BB} {\bm \cdot} \nab) \mu - \mathcal{D}_5 \, \nabla^4 \mu_5
  \nonumber \\ &&
  + \lambda \, \eta \, \left[{\BB} {\bm \cdot} (\nab   \times   {\BB}) - \mu_5 {\BB}^2 \right] ,
\label{mu5-DNS_CSE}
\end{eqnarray}
where $\BB$ and $\UU$ are the magnetic and velocity fields, respectively, 
$\eta$ is the microscopic magnetic diffusivity, 
$p$ is the pressure, $\nu$ is the viscosity, $\rho$ is the mass density, $\SSSS$ is the trace-free strain tensor with components
${\sf S}_{ij}=(\partial_j U_i+\partial_i U_j)/2 -
\delta_{ij} ({\bm \nabla}{\bm \cdot} \UU)/3$, and 
$\lambda=3 \hbar c (8 \alphaem / \kB T)^2$
is the chiral feedback parameter.
Here, $T$ is the temperature, $\kB$ is the Boltzmann constant, $c$ is the speed of light, 
$\alphaem \approx 1/137$ is the fine structure constant,
and $\hbar$ is the reduced Planck constant.
Equations~(\ref{ind-DNS_CSE})--(\ref{mu5-DNS_CSE})
are written in Gaussian units and the chemical potentials have been multiplied by a factor $4\alphaem/(\hbar c)$ such that they have 
units of an inverse length \cite{SBR20}.
The chiral separation effect is described by 
the second terms on the RHS of
Eqs.~(\ref{mu-DNS_CSE})--(\ref{mu5-DNS_CSE}).
The coupling between $\mu_5$ and $\mu$, determined by the value of the constants $C_5$ and $C_\mu$, leads to 
chiral magnetic waves (CMWs) 
with the frequency
$\omega_{\rm CMW} \approx  \pm (C_5 \,C_\mu)^{1/2}
k_\mu |\BB|$ \citep{KY11}, 
where $k_\mu$ is the inverse length scale 
over which $\mu$ changes 
along the magnetic field $\BB$.
For numerical stability, the evolution equations for 
$\mu_5$ and $\mu$ also include 
artificial (hyper-) diffusion terms with 
small diffusion coefficients
$\mathcal{D}_5$ and $\mathcal{D}_\mu$, respectively \citep{SchoberEtAl2022b}. 

For our numerical analysis, we use the \textsc{Pencil Code} \cite{PencilCode2021} 
and solve Eqs.~(\ref{ind-DNS_CSE})--(\ref{mu5-DNS_CSE})
in a three-dimensional periodic domain of size $L^3$ 
with a resolution of up to $N^3=1024^3$ mesh points.
This code employs a third-order accurate time-stepping method \cite{Wil80}
and sixth-order explicit finite differences in space \cite{BD02,Bra03}.
The numerical domain ranges from the minimum wave number $k_1 = 2\pi/L$ 
up to the Nyquist wave number $k_\mathrm{Ny} = N k_1/2$.
We use an isothermal equation of state with $p=\rho\cs^2$, 
where $\cs$ is the sound speed.
The density is initially uniform and, because of mass 
conservation, its value is always equal to the mean 
density $\meanrho$, where angle brackets denote averaging.
In the following, we set $k_1 = \cs = \meanrho = 1$.
For the magnetic Prandtl number, ${\rm Pr}_{\rm M}=\nu/\eta$, we choose
${\rm Pr}_{\rm M}=1$, and $\mathcal{D}_5$ and $\mathcal{D}_\mu$ are chosen
such that the dissipation rates of $\mu_5$ and $\mu$ equal the ones of $\BB$ and $\UU$ 
at $k=k_\mathrm{Ny}$. 
In our analysis, time is either normalized by the diffusion 
time, $t_\eta = (\eta k_1^2)^{-1}=\eta^{-1}$,
or by the period of the CMW, $t_\mathrm{CMW} = 2 \pi / \omega_\mathrm{CMW}$.

\begin{table}
\small
\centering
\caption{Summary of the simulations.}
\begin{tabular}{ l | l | ll | llll  }
\hline
Run   &  Res.    & $C_5 = C_\mu$   & $\; \; \; \lambda$  &  \multicolumn{4}{l}{maximum~value of} \\
~     &  ~       & ~               & ~          &  $\mu_{\mathrm{5,max}}$   &  $B_\mathrm{rms}$  & $\Rm$ & $\mathrm{Lu}$  \\
\hline
$R1$    &     $1024^3$   &  $3$     &  $4 \times 
10^{2}$  &  $68$ & $0.12$ & $77$  &  $440$ \\
$R2$    &     $672^3$   &  $10$    & $4 \times 10^{4}$   & $38$  & $0.024$ & $36$  &  $98$  \\
\hline
\end{tabular}
\label{tab_DNSoverview}
\end{table}
\begin{figure}
\centering
  \includegraphics[width=0.5\textwidth]{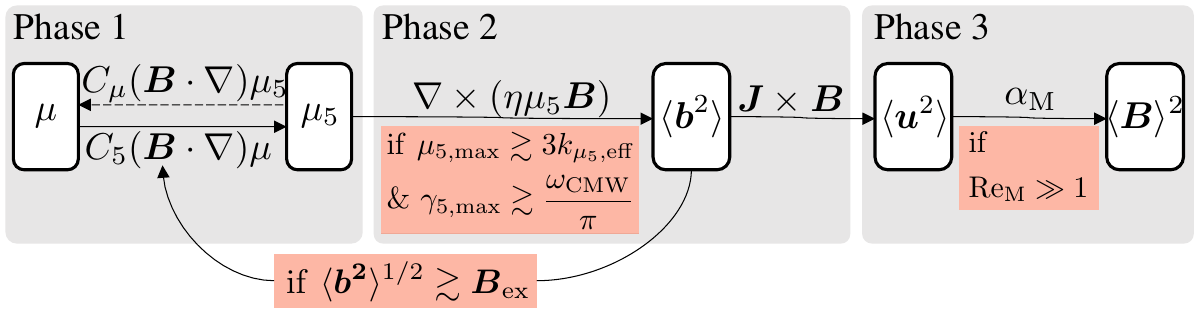} 
  \caption{Illustration of the dynamics in a plasma 
  with autonomous production of chiral asymmetry (Phase 1), accompanied by a generation of fluctuations in the magnetic field
   $\bra{\boldsymbol{b}^2}^{1/2}$
  (Phase 2) and the velocity field $\bra{\boldsymbol{u}^2}^{1/2}$.
  If the magnetic Reynolds number
  becomes larger than unity, 
  the magnetic $\alpha$ effect produces a large-scale magnetic field
  $\bra{\boldsymbol{B}}$ (Phase 3).}
  \label{fig_sketch}
\end{figure}

Fluctuations of $\mu$ at the initial time $t_0$ are
chosen as Gaussian noise with a
power spectrum $E_\mu(k)$, i.e.,
$E_\mu(k,t_0) = E_{\mu,0} k^{s}$.
Here, $E_\mu(k)$ is normalized such that
$\int E_\mu(k)\,\mathrm{d}k=\bra{\mu^2}$.
The amplitude $E_{\mu,0}$ is chosen such that
the maximum values of $\mu$ are comparable
for all runs at the time of the onset of
the small-scale chiral instability.
To allow for chiral magnetic waves 
in the simulations, 
we apply 
a small external magnetic field $\BB_\mathrm{ex}=(B_\mathrm{ex}, 0, 0)$, 
which effectively produces the chiral anomaly.
We further consider a zero initial velocity field $\UU$
and weak perturbations of the initial magnetic field $\BB$ in the form of Gaussian noise.
A summary of runs discussed in this Letter is given in Table~\ref{tab_DNSoverview},
where ${\rm Re}_{\rm M}=U_{\rm rms}/(k_{\rm int}\eta)$ 
is the magnetic Reynolds number, 
${\rm Lu}=U_{\rm A, rms}/(k_{\rm int}\eta)$ is the Lundquist number, 
$U_{\rm rms}$ is the rms velocity fluctuations, and
$U_{\rm A, rms}$ is the Alfv\'en speed based on the rms magnetic fluctuations. 
Further, 
$k_{\rm int}^{-1} = {\cal E}_{\rm M}^{-1}\int k^{-1} E_\mathrm{M}(k)\,\mathrm{d}k$
is the integral scale of the magnetically driven turbulence,
where $E_\mathrm{M}(k)$ is the magnetic energy spectrum
with its peak close to $k_{\rm int}$ and
${\cal E}_{\rm M}= \bra{\BB^2}/2 = \int E_\mathrm{M}(k)\,\mathrm{d}k$ is 
the turbulent magnetic energy density.

Whether enough chiral asymmetry
can be produced to trigger 
the chiral dynamo instability depends on the characteristic
time scales of the system \cite{SchoberEtAl2023b}. 
In particular, the chiral dynamo instability
needs to occur on a time scale shorter than 
half the period $t_\mathrm{CMW}$ of the CMW, $t_\mathrm{CMW}/2=\pi/\omega_\mathrm{CMW}$.
Otherwise, when $t>t_\mathrm{CMW}/2$, the chiral chemical potential 
changes its sign in the CMW, and 
the chiral dynamo instability starts again.
Here, for a linear CMW, the maximum value that $\mu_{5}$ can reach 
is approximately the
initial value of $\mu$, i.e., $\mathrm{max}(\mu_{5,\mathrm{max}})=\mu_{\rm max}(t_0)$.
In this Letter, we focus on times $t < t_\mathrm{CMW}$.


\begin{figure}[t!]
\centering
  \includegraphics[width=0.5\textwidth]{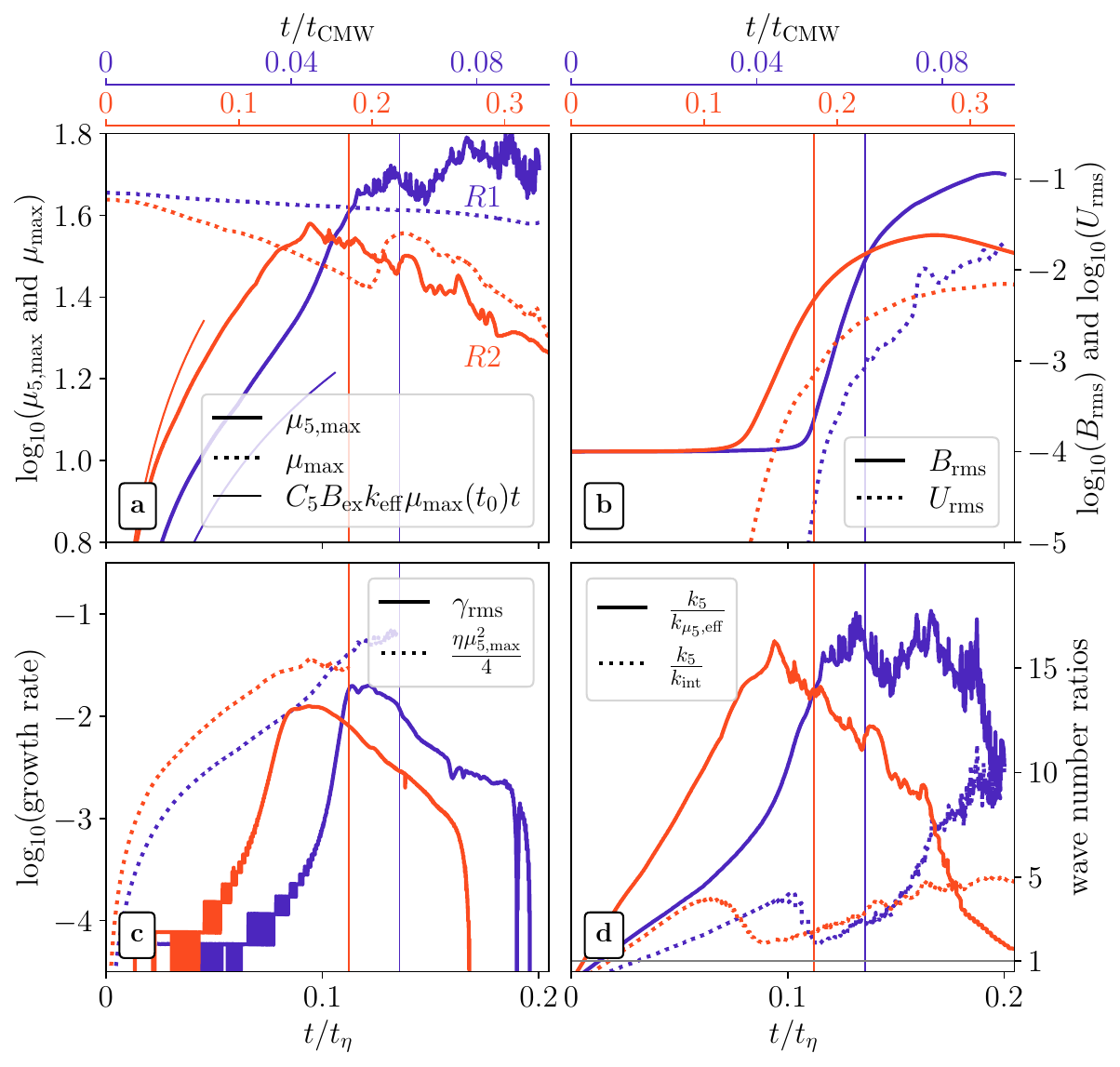} 
  \caption{Comparison of 
  Runs $R1$ (blues lines) and $R2$ (red lines).
  \textit{(a)} Time evolution of the maximum values of 
  $\mu_5$ (solid lines) and 
  $\mu$ (dotted lines).
  \textit{(b)} Time evolution of $B_\mathrm{rms}$ (solid lines) and $U_\mathrm{rms}$ (dotted lines).
  \textit{(c)} Time evolution of the measured growth rate of $B_\mathrm{rms}$, $\gamma_\mathrm{rms}$ (solid lines), and comparison with the theoretical expectation for the chiral dynamo instability (dotted lines).
  \textit{(d)} 
  Time evolution of the ratio of the wave number $k_5$, on which the chiral dynamo instability occurs,
  over the effective correlation 
  wave number $k_{\mu_5,\mathrm{eff}}$ of $\mu_5$.
  Thin vertical lines indicate the respective times 
  when the magnetic Reynolds number ${\rm Re}_{\rm M}$ becomes larger than unity.
  }
  \label{fig_ts_m2_turb}
\end{figure}

\begin{figure}[t!]
\centering
  \includegraphics[width=0.5\textwidth]{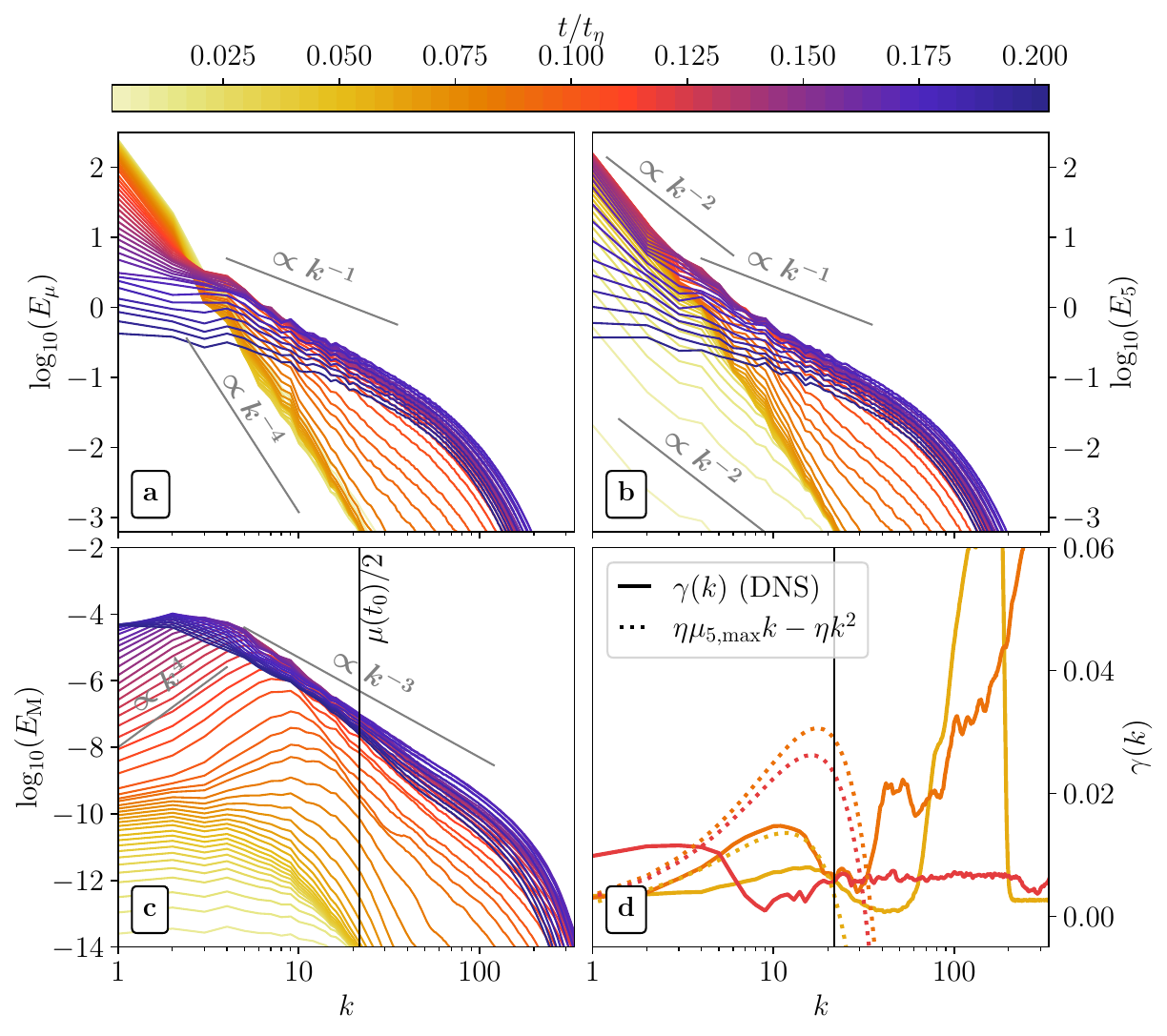} 
  \caption{Evolution of spectra for Run $R2$ with the time indicated by
  the colorbar. 
  \textit{(a)} 
  Spectrum $E_\mu(k)$ of $\mu$.
  \textit{(b)} 
   Spectrum $E_5(k)$ of $\mu_5$.
  \textit{(c)} Magnetic energy spectrum 
  $E_\mathrm{M}(k)$.
  \textit{(d)} Growth rate of the magnetic field strength as a function of 
  wave number, $\gamma(k)$ (solid lines)
  for three different times ($t/t_\eta=0.06$, $0.09$, and $0.12$), and
  theoretically expected growth rate 
  (dotted lines).
  The thin vertical lines indicate the wave number of the chiral dynamo instability if the initial $\mu$ were completely converted into $\mu_5$.}
  \label{fig_spec_gamma}
\end{figure}

The different stages of chiral asymmetry production and magnetic field amplification 
are shown in Fig.~\ref{fig_sketch}.
In Phase~1, a chiral asymmetry 
is produced
via the second term on the RHS in Eq.~(\ref{mu5-DNS_CSE}).
For times less than the period of the CMW, i.e., 
$t \ll t_\mathrm{CMW}$
and as long as $B_\mathrm{ex}$ dominates over magnetic fluctuations, the evolution of $\mu_5$ is described as
\begin{eqnarray}
  |\mu_5(t)| 
  \approx C_5 |({\BB} {\bm \cdot} \nab)  \mu(t) |\, t 
  \approx C_5 |B_\mathrm{ex}  k_{\mu,\mathrm{eff}}(t_0) \mu(t_0) |\, t,
\label{eq_mu5_init_gen}
\end{eqnarray}
where $k_{\mu,\mathrm{eff}}$ is the characteristic wave number of the initial $\mu$.
Introducing the spectrum of the chiral chemical 
potential $E_5(k)$ as $\int E_5(k)\,\mathrm{d}k=\bra{\mu_5^2}$,
and using the estimate for $|\mu_5(t)|$ given by 
Eq.~(\ref{eq_mu5_init_gen}), we relate the spectrum functions $E_5(k)$ and $E_\mu(k)$ as
$\left[E_5(k) k\right]^{1/2}  \approx C_5  B_\mathrm{ex} k  \left[E_\mu(k) k\right]^{1/2} \, t$.
Hence, initial spatial fluctuations of the chemical potential with a spectrum $E_\mu(k) \propto k^{s}$ produce fluctuations of the chiral chemical potential with the spectrum
\begin{eqnarray}
  E_5(k) \propto k^{2+s} ~t^2.
\label{eq_E_5k}
\end{eqnarray}
The time evolution determined by Eq.~(\ref{eq_mu5_init_gen}) 
is seen in the initial times of Runs $R1$ and $R2$
(see Fig.~\ref{fig_ts_m2_turb}a).
Moreover, the obtained spectrum $E_5(k)$ in 
Phase~1 of Run $R2$ is $\propto k^{-2}$ 
(see Fig.~\ref{fig_spec_gamma}b), as expected from 
Eq.~(\ref{eq_E_5k}), 
where the initial spectrum of the chemical potential
has the exponent $s =-4$; see Fig.~\ref{fig_spec_gamma}a.

Phase~2 starts when $\mu_5$ exceeds a critical value
for the excitation of the chiral dynamo instability,
resulting in exponential growth of magnetic fluctuations.
The growth rate of this instability is estimated as
\begin{eqnarray}
   \gamma(k) \approx \eta \mu_{5,\mathrm{max}} k - \eta k^2,
\label{eq_gamma}
\end{eqnarray}
where $\mu_{5,\mathrm{max}}$ is the spatial maximum of $\mu_5$
\citep{SchoberEtAl2022a,SchoberEtAl2022b}. 
The maximum possible growth rate is 
\begin{eqnarray}
   \gamma_5(t)= \eta \mu_{5,\mathrm{max}}^2(t)/4,
\label{eq_gammak}
\end{eqnarray}
but it is only reached when the instability wave number
$k_5(t)= \mu_{5,\mathrm{max}}(t)/2$,
is much larger than the effective correlation 
wave number of the fluctuations of $\mu_5$
\begin{eqnarray}
  k_{\mu_5,\mathrm{eff}}^{-1}(t) = 
  \bra{\mu_5^2}^{-1} 
  \int  k^{-1} E_5(k)\,\mathrm{d}k.
\label{eq_kmu5eff}
\end{eqnarray}
The latter condition applies only to systems
in which the initial $\mu$ or $\mu_5$ are dominated by fluctuations
\citep{SchoberEtAl2022a,SchoberEtAl2022b}.

The time evolution of the rms magnetic field ($B_\mathrm{rms}$), the 
measured growth rate of $B_\mathrm{rms}$ ($\gamma_\mathrm{rms}$), and 
the ratio $k_5/k_{\mu_5,\mathrm{eff}}$ are shown 
in Fig.~\ref{fig_ts_m2_turb} for runs $R1$ and $R2$.
We find that the maximum growth rate of the chiral dynamo instability is
attained when the scale separation ratio obeys 
$k_5/k_{\mu_5,\mathrm{eff}} \gtrsim 10$,
as can be seen when comparing Figs.~\ref{fig_ts_m2_turb}c and~\ref{fig_ts_m2_turb}d.
In Fig.~\ref{fig_spec_gamma}d, we show the measured growth rate of the scale-dependent magnetic field strength and compare it 
with the rough estimate for the
growth rate given by Eq.~(\ref{eq_gammak})
for three instants during Phase~2.
While the measured curves $\gamma(k)$ do have a peak at 
higher $k$ initially, they become more and more flat
with time, indicating efficient mode coupling.
Such a $k$-independent growth rate has also been observed in simulations
of chiral magnetohydrodynamics (MHD) with a spatially homogeneous $\mu_5$
when turbulence is driven by an artificial forcing
term in the Navier-Stokes equation \cite{SchoberEtAl2017}
and in the kinematic stage of a helically driven large-scale dynamo in classical MHD \cite{SubramanianBrandenburg2014}.
Note, that during Phase~2, $\mu_5$ continues to grow,
and the field $B_{\mathrm{ex}}$ is being replaced by $B_{\mathrm{ex}}+b(t)$, once $b(t)\gtrsim B_{\mathrm{ex}}$, where $b$ are magnetic fluctuations.
Therefore, the $\mu_5$ production phase becomes nonlinear. 
For Run $R1$, 
the transition to the nonlinear chiral asymmetry 
production phase occurs at $t\approx 0.1\,t_\eta$; 
see Fig.~\ref{fig_ts_m2_turb}a.

\begin{figure}[t!]
\centering
  \includegraphics[width=0.5\textwidth]{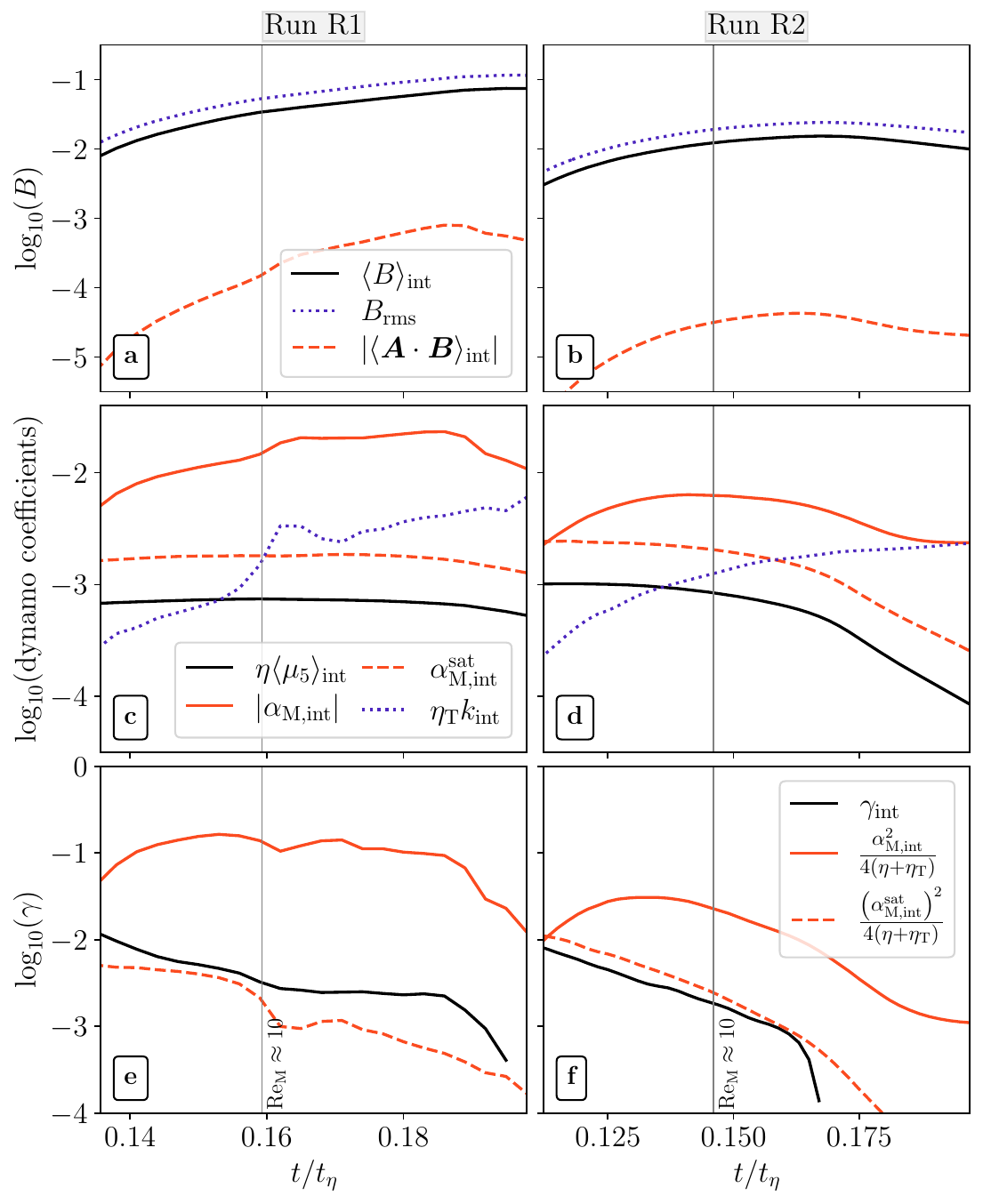} 
  \caption{Mean-field analysis for Runs $R1$ (left panels) and $R2$ (right panels).
  \textit{(a)} and \textit{(b)} Time evolution of the mean magnetic field
  $\bra{B}_{\mathrm{int}}$ (black lines),
  $B_\mathrm{rms}$ (dotted blue lines), 
  magnetic helicity 
  $\bra{\boldsymbol{A}\cdot \boldsymbol{B}}_{\mathrm{int}}$ (dashed red lines).
  \textit{(c)} and \textit{(d)} Time evolution of 
  $\eta \bra{\mu_5}_{\mathrm{int}}$ (black lines),
  $\alpha_\mathrm{M, \mathrm{int}}$ (red lines) and 
  $\eta_\mathrm{T} k_\mathrm{int}$ (dotted blue lines).
  \textit{(e)} and \textit{(f)}  
  Time evolution of the growth rate $\gamma_\mathrm{int}$ (black solid line) of 
  $\bra{B}_{\mathrm{int}}$
  compared to the mean-field dynamo prediction, where the maximum growth rate is based on
  $\alpha_\mathrm{M,int}$ (solid red line) and $\alpha_\mathrm{M,int}^{\rm sat}$
  (dashed red line).
  The time evolution presented here starts at the time 
  when $\Rm$ exceeds unity and 
  thin vertical lines indicate when $\Rm$ becomes larger than $10$.
  }
  \label{fig_ts_m2_meanfield}
\end{figure}

Eventually, the chiral dynamo instability 
reaches its nonlinear stage since the Lorentz force 
[the first term on the RHS of Eq.~(\ref{UU-DNS_CSE})] 
increases with time due to the nonlinear evolution of the 
magnetic field. 
The Lorentz force produces velocity fluctuations
(see Fig.~\ref{fig_ts_m2_turb}b),
and the fluid and magnetic Reynolds numbers 
become larger than unity. 
Thus, Phase~3 begins, magnetically dominated turbulence is produced, 
and a large-scale magnetic field is generated
via a mean-field dynamo instability,
which is excited with the growth rate
\begin{eqnarray}
\gamma_\alpha=\left(\eta \langle{\mu}_{5}\rangle + \alpha_\mathrm{M}\right) k - (\eta+\eta_\mathrm{T}) k^2.
\label{Eq10a}
\end{eqnarray}
Here, $\alpha_\mathrm{M} = 2(q-1)/(q+1) \, 
\tau_{\rm c} \, \chi_{\rm c}$ 
is the magnetic $\alpha$ effect, which is determined by the current helicity
$\chi_{\rm c} = \bra{{\bm b} {\bm \cdot} ({\bm \nabla} \times{\bm b})} \approx \bra{ {\bm a}\cdot {\bm b}} \, k_\mathrm{int}^2$,  
where $q$ is the exponent of the magnetic energy spectrum $E_{\rm M}\propto k^{-q}$, 
and $\bra{\mu_5}$ is the mean chiral chemical potential.
As follows from Fig.~\ref{fig_spec_gamma}c, 
the exponent is $q\approx3$.
The correlation time of the magnetically driven turbulence is $\tau_{\rm c} \approx (U_\mathrm{A} k_\mathrm{int})^{-1}$, where the Alfv\'en speed is
$U_{\rm A}=\sqrt{\bra{{\bm b}^2}}
\approx B_\mathrm{rms}$. 
The mean fluid density $\meanrho$ entering in $U_{\rm A}$ and $\alpha_\mathrm{M}$ is set to unity.
The turbulent diffusion coefficient $\eta_\mathrm{T}$ is estimated as 
$\eta_\mathrm{T}=U_\mathrm{rms}/(3 k_\mathrm{int})$.
The growth rate $\gamma_\alpha$ of the mean-field dynamo
instability attains the maximum value
\begin{eqnarray}
  \gamma_\alpha^{\rm max}= \frac{(\eta 
  \langle{\mu}_{5}\rangle  + \alpha_\mathrm{M})^2}{4(\eta+\eta_\mathrm{T})} .
\label{Eq10b}
\end{eqnarray}
The current helicity $\chi_{\rm c}$ and magnetic $\alpha$ effect 
$\alpha_\mathrm{M}$ can also be estimated 
from the evolutionary equation for the magnetic helicity $\bra{ {\bm a}\cdot {\bm b}}$ of the small-scale field ${\bm b}={\bm \nabla} \times{\bm a}$ in chiral MHD \citep{RogachevskiiEtAl2017}:
\begin{eqnarray}
{\partial \over \partial t}  \bra{{\bm a} {\bm \cdot} {\bm b}} + 
\nabla \cdot {\bm F}
&=& 2 \eta \, \meanmufive \bra{{\bm b}^2} - 2 \meanEMF \cdot \meanBB 
\nonumber\\
&&- 2 \eta \, \bra{{\bm b} \, ({\bm \nabla} \times {\bm b})} ,
\label{MH1}
\end{eqnarray}
where ${\bm F}$ is the flux of $\bra{{\bm a} {\bm \cdot} {\bm b}}$
and $\meanEMF \equiv \bra{{\bm u} {\bm \times} {\bm b}}=\alpha_\mathrm{M} \meanBB - \eta_T \, ({\bm \nabla} \times \meanBB)$
is the turbulent electromotive force.
In the steady-state, two leading source or sink
terms in Eq.~(\ref{MH1}),
$2 \eta \, \meanmufive  \bra{{\bm b}^2} - 2 \alpha_\mathrm{M}  \meanBB^2$,
compensate each other, so that
the magnetic $\alpha$ effect reaches
$\alpha_\mathrm{M}^{\rm sat} = \eta \, \meanmufive \, {\bra{{\bm b}^2} / \meanBB^2}$
\citep{SchoberEtAl2022a,SchoberEtAl2022b}.

We present an analysis of the mean-field dynamo stage
for Runs $R1$ and $R2$ in Fig.~\ref{fig_ts_m2_meanfield}. 
In our system, where velocity fluctuations are driven by an increasing Lorentz force during the nonlinear phase of the
chiral dynamo instability, the integral scale
of turbulence, $k_{\rm int}^{-1}$, increases with time.
This implies that the turbulent range of scales is expanding in time. 
To take this fact into account, we perform averaging over 
the scales larger than $k_\mathrm{int}^{-1}$
at a given time.
In particular, the
mean quantities used in Fig.~\ref{fig_ts_m2_meanfield} are calculated in the simulations as
$\bra{B}_{\mathrm{int}} \equiv \left[2 \int E_\mathrm{M}(k) f(k)\,\mathrm{d}k\right]^{1/2}$,
$\bra{\mu_5}_{\mathrm{int}} \equiv \left[\int E_5(k)f(k) \,\mathrm{d}k\right]^{1/2}$,
and 
$\bra{{\bm A} {\bm \cdot} {\bm B}}_\mathrm{int}\equiv \int H_\mathrm{M}(k) f(k) \,\mathrm{d}k$,
where $f(k)\equiv\left[1-\mathrm{tanh}(k-k_\mathrm{int})\right]/2$, and $H_\mathrm{M}(k)$ is the magnetic helicity spectrum. 
From these averages we calculate 
$\alpha_\mathrm{M, int} = \bra{{\bm A} {\bm \cdot} {\bm B}}_\mathrm{int} k_\mathrm{int}/B_\mathrm{rms}$  
and $\alpha_\mathrm{M, int}^{\rm sat} = \eta \bra{\mu_5}_\mathrm{int} B_\mathrm{rms}^2/\bra{B}_{\mathrm{int}}^2$. 
As can be seen in panels (e) and (f) of 
Fig.~\ref{fig_ts_m2_meanfield}, the measured growth
rate of the mean magnetic field is comparable with the 
theoretical predictions based on the magnetic $\alpha$ effect.
In the mean-field dynamo phase, the dynamo generation term 
$\alpha_\mathrm{M, int}^{\rm sat} k_\mathrm{int}$ is larger 
than the damping term $\eta_\mathrm{T} k_\mathrm{int}^2$.
However, since the level of turbulence increases,
the turbulent diffusion coefficient $\eta_\mathrm{T}$ increases
and $\alpha_\mathrm{M, int}^{\rm sat}$ decreases,
leading finally to the end of the mean-field dynamo instability.

Since $\alpha_\mathrm{M}$ and $\eta_\mathrm{T}$ change in time in the simulations, Eqs.~(\ref{Eq10a}) and~(\ref{Eq10b}) serve only as
rough estimates on the order of magnitude for the growth rates of the mean-field dynamo instability used for comparisons with DNS.
A similar statement is also valid for 
Eqs.~(\ref{eq_gamma})--(\ref{eq_gammak}) for the growth
rates of the chiral dynamo instability used for comparisons with DNS.

The dynamo instability can,
in principle, be followed by an inverse cascade of 
magnetic energy that is related to a conservation law. 
In chiral MHD, the total chirality
(the sum of the chiral chemical potential and
the magnetic helicity multiplied by $\lambda/2$)
is conserved. 
In this study, both the initial magnetic helicity
and the chiral chemical potential vanish, so the 
anomaly-induced inverse cascade related to the 
decrease of chiral chemical potential due to
the increase of magnetic helicity during the nonlinear 
chiral dynamo (as described in \cite{BFR12}) is not present.
However, it has been shown that in chiral MHD 
with vanishing total chirality an adaptation of 
the Hosking integral \cite{HS21} is conserved \cite{BKS2023}, which leads to 
an increase of the magnetic correlation length during the decay of 
magnetic energy.
In our present study, such a long-term evolution cannot be studied because
at dynamo saturation the magnetic correlation length has already reached the size of the numerical domain.


In conclusion, we have found a scenario in which a chiral asymmetry is produced 
in a plasma with initially balanced numbers of left- and
right-handed electrically charged fermions due to the joint action of
the chiral separation effect and
spatial inhomogeneity 
in fluctuations of the chemical potential.
This causes a chiral dynamo instability 
and the subsequent production of magnetically driven turbulence followed by
large-scale magnetic field generation through the magnetic $\alpha$ effect.
This scenario can be relevant for the early Universe if chirality is not produced by other mechanisms. 
It might be less relevant for proto-neutron stars 
since a net chiral chemical potential can there be produced
due to neutrino emission \cite{SiglLeite2016}.
Beyond astrophysics, this scenario might also be relevant for condensed matter systems.

\begin{acknowledgements}
This study was initiated several years ago 
through productive discussions with Dmitry Kharzeev.
J.S.~acknowledges the support from the Swiss National Science Foundation under Grant No.\ 185863.
A.B.~was supported in part through a grant from the Swedish Research Council
(Vetenskapsr{\aa}det, 2019-04234).
Nordita was sponsored by NordForsk.
We acknowledge the allocation of computing resources provided by the
Swedish National Allocations Committee at the Center for
Parallel Computers at the Royal Institute of Technology in Stockholm.
\end{acknowledgements}


\begin{thebibliography}{42}%
\makeatletter
\providecommand \@ifxundefined [1]{%
 \@ifx{#1\undefined}
}%
\providecommand \@ifnum [1]{%
 \ifnum #1\expandafter \@firstoftwo
 \else \expandafter \@secondoftwo
 \fi
}%
\providecommand \@ifx [1]{%
 \ifx #1\expandafter \@firstoftwo
 \else \expandafter \@secondoftwo
 \fi
}%
\providecommand \natexlab [1]{#1}%
\providecommand \enquote  [1]{``#1''}%
\providecommand \bibnamefont  [1]{#1}%
\providecommand \bibfnamefont [1]{#1}%
\providecommand \citenamefont [1]{#1}%
\providecommand \href@noop [0]{\@secondoftwo}%
\providecommand \href [0]{\begingroup \@sanitize@url \@href}%
\providecommand \@href[1]{\@@startlink{#1}\@@href}%
\providecommand \@@href[1]{\endgroup#1\@@endlink}%
\providecommand \@sanitize@url [0]{\catcode `\\12\catcode `\$12\catcode
  `\&12\catcode `\#12\catcode `\^12\catcode `\_12\catcode `\%12\relax}%
\providecommand \@@startlink[1]{}%
\providecommand \@@endlink[0]{}%
\providecommand \url  [0]{\begingroup\@sanitize@url \@url }%
\providecommand \@url [1]{\endgroup\@href {#1}{\urlprefix }}%
\providecommand \urlprefix  [0]{URL }%
\providecommand \Eprint [0]{\href }%
\providecommand \doibase [0]{https://doi.org/}%
\providecommand \selectlanguage [0]{\@gobble}%
\providecommand \bibinfo  [0]{\@secondoftwo}%
\providecommand \bibfield  [0]{\@secondoftwo}%
\providecommand \translation [1]{[#1]}%
\providecommand \BibitemOpen [0]{}%
\providecommand \bibitemStop [0]{}%
\providecommand \bibitemNoStop [0]{.\EOS\space}%
\providecommand \EOS [0]{\spacefactor3000\relax}%
\providecommand \BibitemShut  [1]{\csname bibitem#1\endcsname}%
\let\auto@bib@innerbib\@empty
\bibitem [{\citenamefont {Vilenkin}(1980)}]{Vilenkin:80a}%
  \BibitemOpen
  \bibfield  {author} {\bibinfo {author} {\bibfnamefont {A.}~\bibnamefont
  {Vilenkin}},\ }\bibfield  {title} {\bibinfo {title} {{Equilibrium parity
  violating current in a magnetic field}},\ }\href@noop {} {\bibfield
  {journal} {\bibinfo  {journal} {Phys.~Rev. D}\ }\textbf {\bibinfo {volume}
  {22}},\ \bibinfo {pages} {3080} (\bibinfo {year} {1980})}\BibitemShut
  {NoStop}%
\bibitem [{\citenamefont {{Redlich}}\ and\ \citenamefont
  {{Wijewardhana}}(1985)}]{RW85}%
  \BibitemOpen
  \bibfield  {author} {\bibinfo {author} {\bibfnamefont {A.~N.}\ \bibnamefont
  {{Redlich}}}\ and\ \bibinfo {author} {\bibfnamefont {L.~C.~R.}\ \bibnamefont
  {{Wijewardhana}}},\ }\bibfield  {title} {\bibinfo {title} {{Induced
  Chern-Simons terms at high temperatures and finite densities}},\ }\href@noop
  {} {\bibfield  {journal} {\bibinfo  {journal} {Phys.~Rev.~Lett.}\ }\textbf
  {\bibinfo {volume} {54}},\ \bibinfo {pages} {970} (\bibinfo {year}
  {1985})}\BibitemShut {NoStop}%
\bibitem [{\citenamefont {Tsokos}(1985)}]{Tsokos:85}%
  \BibitemOpen
  \bibfield  {author} {\bibinfo {author} {\bibfnamefont {K.}~\bibnamefont
  {Tsokos}},\ }\bibfield  {title} {\bibinfo {title} {{Topological mass terms
  and the high temperature limit of chiral gauge theories}},\ }\href@noop {}
  {\bibfield  {journal} {\bibinfo  {journal} {Phys.~Lett.~B}\ }\textbf
  {\bibinfo {volume} {157}},\ \bibinfo {pages} {413} (\bibinfo {year}
  {1985})}\BibitemShut {NoStop}%
\bibitem [{\citenamefont {Fr\"{o}hlich}\ and\ \citenamefont
  {Pedrini}(2000)}]{Frohlich:2000en}%
  \BibitemOpen
  \bibfield  {author} {\bibinfo {author} {\bibfnamefont {J.}~\bibnamefont
  {Fr\"{o}hlich}}\ and\ \bibinfo {author} {\bibfnamefont {B.}~\bibnamefont
  {Pedrini}},\ }\bibfield  {title} {\bibinfo {title} {{New applications of the
  chiral anomaly}},\ }in\ \href@noop {} {\emph {\bibinfo {booktitle}
  {Mathematical Physics 2000}}},\ \bibinfo {series and number} {International
  Conference on Mathematical Physics 2000, Imperial college (London)},\
  \bibinfo {editor} {edited by\ \bibinfo {editor} {\bibfnamefont {A.~S.}\
  \bibnamefont {Fokas}}, \bibinfo {editor} {\bibfnamefont {A.}~\bibnamefont
  {Grigoryan}}, \bibinfo {editor} {\bibfnamefont {T.}~\bibnamefont {Kibble}},\
  and\ \bibinfo {editor} {\bibfnamefont {B.}~\bibnamefont {Zegarlinski}}}\
  (\bibinfo  {publisher} {World Scientific Publishing Company},\ \bibinfo
  {year} {2000})\BibitemShut {NoStop}%
\bibitem [{\citenamefont {Fr\"{o}hlich}\ and\ \citenamefont
  {Pedrini}(2002)}]{Frohlich:2002fg}%
  \BibitemOpen
  \bibfield  {author} {\bibinfo {author} {\bibfnamefont {J.}~\bibnamefont
  {Fr\"{o}hlich}}\ and\ \bibinfo {author} {\bibfnamefont {B.}~\bibnamefont
  {Pedrini}},\ }\bibfield  {title} {\bibinfo {title} {{Axions, quantum
  mechanical pumping, and primeval magnetic fields}},\ }in\ \href@noop {}
  {\emph {\bibinfo {booktitle} {Statistical Field Theory}}},\ \bibinfo {editor}
  {edited by\ \bibinfo {editor} {\bibfnamefont {A.}~\bibnamefont {Cappelli}}\
  and\ \bibinfo {editor} {\bibfnamefont {G.}~\bibnamefont {Mussardo}}}\
  (\bibinfo  {publisher} {Kluwer},\ \bibinfo {year} {2002})\BibitemShut
  {NoStop}%
\bibitem [{\citenamefont {Fukushima}\ \emph {et~al.}(2008)\citenamefont
  {Fukushima}, \citenamefont {Kharzeev},\ and\ \citenamefont
  {Warringa}}]{Fukushima:08}%
  \BibitemOpen
  \bibfield  {author} {\bibinfo {author} {\bibfnamefont {K.}~\bibnamefont
  {Fukushima}}, \bibinfo {author} {\bibfnamefont {D.~E.}\ \bibnamefont
  {Kharzeev}},\ and\ \bibinfo {author} {\bibfnamefont {H.~J.}\ \bibnamefont
  {Warringa}},\ }\bibfield  {title} {\bibinfo {title} {{The Chiral Magnetic
  Effect}},\ }\href@noop {} {\bibfield  {journal} {\bibinfo  {journal}
  {Phys.~Rev. D}\ }\textbf {\bibinfo {volume} {78}},\ \bibinfo {pages} {074033}
  (\bibinfo {year} {2008})}\BibitemShut {NoStop}%
\bibitem [{\citenamefont {Son}\ and\ \citenamefont
  {Surowka}(2009)}]{Son:2009tf}%
  \BibitemOpen
  \bibfield  {author} {\bibinfo {author} {\bibfnamefont {D.~T.}\ \bibnamefont
  {Son}}\ and\ \bibinfo {author} {\bibfnamefont {P.}~\bibnamefont {Surowka}},\
  }\bibfield  {title} {\bibinfo {title} {{Hydrodynamics with Triangle
  Anomalies}},\ }\href@noop {} {\bibfield  {journal} {\bibinfo  {journal}
  {Phys.~Rev.~Lett.}\ }\textbf {\bibinfo {volume} {103}},\ \bibinfo {pages}
  {191601} (\bibinfo {year} {2009})}\BibitemShut {NoStop}%
\bibitem [{\citenamefont {{Alekseev}}\ \emph {et~al.}(1998)\citenamefont
  {{Alekseev}}, \citenamefont {{Cheianov}},\ and\ \citenamefont
  {{Fr{\"o}hlich}}}]{AlekseevEtAl1998}%
  \BibitemOpen
  \bibfield  {author} {\bibinfo {author} {\bibfnamefont {A.~Y.}\ \bibnamefont
  {{Alekseev}}}, \bibinfo {author} {\bibfnamefont {V.~V.}\ \bibnamefont
  {{Cheianov}}},\ and\ \bibinfo {author} {\bibfnamefont {J.}~\bibnamefont
  {{Fr{\"o}hlich}}},\ }\bibfield  {title} {\bibinfo {title} {{Universality of
  transport properties in equilibrium, the Goldstone theorem, and chiral
  anomaly}},\ }\href@noop {} {\bibfield  {journal} {\bibinfo  {journal}
  {Phys.~Rev.~Lett.}\ }\textbf {\bibinfo {volume} {81}},\ \bibinfo {pages}
  {3503} (\bibinfo {year} {1998})}\BibitemShut {NoStop}%
\bibitem [{\citenamefont {Giovannini}(2013)}]{GI13}%
  \BibitemOpen
  \bibfield  {author} {\bibinfo {author} {\bibfnamefont {M.}~\bibnamefont
  {Giovannini}},\ }\bibfield  {title} {\bibinfo {title} {Anomalous
  magnetohydrodynamics},\ }\href@noop {} {\bibfield  {journal} {\bibinfo
  {journal} {Phys. Rev. D}\ }\textbf {\bibinfo {volume} {88}},\ \bibinfo
  {pages} {063536} (\bibinfo {year} {2013})}\BibitemShut {NoStop}%
\bibitem [{\citenamefont {Rogachevskii}\ \emph {et~al.}(2017)\citenamefont
  {Rogachevskii}, \citenamefont {Ruchayskiy}, \citenamefont {Boyarsky},
  \citenamefont {Fr\"ohlich}, \citenamefont {Kleeorin}, \citenamefont
  {Brandenburg},\ and\ \citenamefont {Schober}}]{RogachevskiiEtAl2017}%
  \BibitemOpen
  \bibfield  {author} {\bibinfo {author} {\bibfnamefont {I.}~\bibnamefont
  {Rogachevskii}}, \bibinfo {author} {\bibfnamefont {O.}~\bibnamefont
  {Ruchayskiy}}, \bibinfo {author} {\bibfnamefont {A.}~\bibnamefont
  {Boyarsky}}, \bibinfo {author} {\bibfnamefont {J.}~\bibnamefont
  {Fr\"ohlich}}, \bibinfo {author} {\bibfnamefont {N.}~\bibnamefont
  {Kleeorin}}, \bibinfo {author} {\bibfnamefont {A.}~\bibnamefont
  {Brandenburg}},\ and\ \bibinfo {author} {\bibfnamefont {J.}~\bibnamefont
  {Schober}},\ }\bibfield  {title} {\bibinfo {title} {{Laminar and Turbulent
  Dynamos in Chiral Magnetohydrodynamics. I. Theory}},\ }\href
  {https://doi.org/10.3847/1538-4357/aa886b} {\bibfield  {journal} {\bibinfo
  {journal} {Astrophys. J.}\ }\textbf {\bibinfo {volume} {846}},\ \bibinfo
  {pages} {153} (\bibinfo {year} {2017})}\BibitemShut {NoStop}%
\bibitem [{\citenamefont {{Hattori}}\ \emph {et~al.}(2019)\citenamefont
  {{Hattori}}, \citenamefont {{Hirono}}, \citenamefont {{Yee}},\ and\
  \citenamefont {{Yin}}}]{HattoriEtAl2019}%
  \BibitemOpen
  \bibfield  {author} {\bibinfo {author} {\bibfnamefont {K.}~\bibnamefont
  {{Hattori}}}, \bibinfo {author} {\bibfnamefont {Y.}~\bibnamefont {{Hirono}}},
  \bibinfo {author} {\bibfnamefont {H.-U.}\ \bibnamefont {{Yee}}},\ and\
  \bibinfo {author} {\bibfnamefont {Y.}~\bibnamefont {{Yin}}},\ }\bibfield
  {title} {\bibinfo {title} {{Magnetohydrodynamics with chiral anomaly: Phases
  of collective excitations and instabilities}},\ }\href
  {https://doi.org/10.1103/PhysRevD.100.065023} {\bibfield  {journal} {\bibinfo
   {journal} {\prd}\ }\textbf {\bibinfo {volume} {100}},\ \bibinfo {eid}
  {065023} (\bibinfo {year} {2019})},\ \Eprint
  {https://arxiv.org/abs/1711.08450} {arXiv:1711.08450 [hep-th]} \BibitemShut
  {NoStop}%
\bibitem [{\citenamefont {{Boyarsky}}\ \emph {et~al.}(2015)\citenamefont
  {{Boyarsky}}, \citenamefont {{Fr{\"o}hlich}},\ and\ \citenamefont
  {{Ruchayskiy}}}]{BFR15}%
  \BibitemOpen
  \bibfield  {author} {\bibinfo {author} {\bibfnamefont {A.}~\bibnamefont
  {{Boyarsky}}}, \bibinfo {author} {\bibfnamefont {J.}~\bibnamefont
  {{Fr{\"o}hlich}}},\ and\ \bibinfo {author} {\bibfnamefont {O.}~\bibnamefont
  {{Ruchayskiy}}},\ }\bibfield  {title} {\bibinfo {title}
  {{Magnetohydrodynamics of chiral relativistic fluids}},\ }\href
  {https://doi.org/10.1103/PhysRevD.92.043004} {\bibfield  {journal} {\bibinfo
  {journal} {\prd}\ }\textbf {\bibinfo {volume} {92}},\ \bibinfo {eid} {043004}
  (\bibinfo {year} {2015})}\BibitemShut {NoStop}%
\bibitem [{\citenamefont {{Grabowska}}\ \emph {et~al.}(2015)\citenamefont
  {{Grabowska}}, \citenamefont {{Kaplan}},\ and\ \citenamefont
  {{Reddy}}}]{GKR15}%
  \BibitemOpen
  \bibfield  {author} {\bibinfo {author} {\bibfnamefont {D.}~\bibnamefont
  {{Grabowska}}}, \bibinfo {author} {\bibfnamefont {D.~B.}\ \bibnamefont
  {{Kaplan}}},\ and\ \bibinfo {author} {\bibfnamefont {S.}~\bibnamefont
  {{Reddy}}},\ }\bibfield  {title} {\bibinfo {title} {{Role of the electron
  mass in damping chiral plasma instability in Supernovae and neutron stars}},\
  }\href {https://doi.org/10.1103/PhysRevD.91.085035} {\bibfield  {journal}
  {\bibinfo  {journal} {\prd}\ }\textbf {\bibinfo {volume} {91}},\ \bibinfo
  {eid} {085035} (\bibinfo {year} {2015})}\BibitemShut {NoStop}%
\bibitem [{\citenamefont {Kharzeev}(2014)}]{KH14}%
  \BibitemOpen
  \bibfield  {author} {\bibinfo {author} {\bibfnamefont {D.~E.}\ \bibnamefont
  {Kharzeev}},\ }\bibfield  {title} {\bibinfo {title} {The chiral magnetic
  effect and anomaly-induced transport},\ }\href@noop {} {\bibfield  {journal}
  {\bibinfo  {journal} {Prog. Part. Nucl. Phys.}\ }\textbf {\bibinfo {volume}
  {75}},\ \bibinfo {pages} {133} (\bibinfo {year} {2014})}\BibitemShut
  {NoStop}%
\bibitem [{\citenamefont {{Kharzeev}}\ \emph {et~al.}(2016)\citenamefont
  {{Kharzeev}}, \citenamefont {{Liao}}, \citenamefont {{Voloshin}},\ and\
  \citenamefont {{Wang}}}]{K16}%
  \BibitemOpen
  \bibfield  {author} {\bibinfo {author} {\bibfnamefont {D.~E.}\ \bibnamefont
  {{Kharzeev}}}, \bibinfo {author} {\bibfnamefont {J.}~\bibnamefont {{Liao}}},
  \bibinfo {author} {\bibfnamefont {S.~A.}\ \bibnamefont {{Voloshin}}},\ and\
  \bibinfo {author} {\bibfnamefont {G.}~\bibnamefont {{Wang}}},\ }\bibfield
  {title} {\bibinfo {title} {{Chiral magnetic and vortical effects in
  high-energy nuclear collisions-A status report}},\ }\href@noop {} {\bibfield
  {journal} {\bibinfo  {journal} {Prog. Part. Nucl. Phys.}\ }\textbf {\bibinfo
  {volume} {88}},\ \bibinfo {pages} {1} (\bibinfo {year} {2016})}\BibitemShut
  {NoStop}%
\bibitem [{\citenamefont {{Schober}}\ \emph {et~al.}(2018)\citenamefont
  {{Schober}}, \citenamefont {{Rogachevskii}}, \citenamefont {{Brandenburg}},
  \citenamefont {{Boyarsky}}, \citenamefont {{Fr{\"o}hlich}}, \citenamefont
  {{Ruchayskiy}},\ and\ \citenamefont {{Kleeorin}}}]{SchoberEtAl2017}%
  \BibitemOpen
  \bibfield  {author} {\bibinfo {author} {\bibfnamefont {J.}~\bibnamefont
  {{Schober}}}, \bibinfo {author} {\bibfnamefont {I.}~\bibnamefont
  {{Rogachevskii}}}, \bibinfo {author} {\bibfnamefont {A.}~\bibnamefont
  {{Brandenburg}}}, \bibinfo {author} {\bibfnamefont {A.}~\bibnamefont
  {{Boyarsky}}}, \bibinfo {author} {\bibfnamefont {J.}~\bibnamefont
  {{Fr{\"o}hlich}}}, \bibinfo {author} {\bibfnamefont {O.}~\bibnamefont
  {{Ruchayskiy}}},\ and\ \bibinfo {author} {\bibfnamefont {N.}~\bibnamefont
  {{Kleeorin}}},\ }\bibfield  {title} {\bibinfo {title} {{Laminar and turbulent
  dynamos in chiral magnetohydrodynamics. II. Simulations}},\ }\href
  {https://doi.org/10.3847/1538-4357/aaba75} {\bibfield  {journal} {\bibinfo
  {journal} {\apj}\ }\textbf {\bibinfo {volume} {858}},\ \bibinfo {eid} {124}
  (\bibinfo {year} {2018})},\ \Eprint {https://arxiv.org/abs/1711.09733}
  {arXiv:1711.09733 [physics.flu-dyn]} \BibitemShut {NoStop}%
\bibitem [{\citenamefont {Schober}\ \emph {et~al.}(2019)\citenamefont
  {Schober}, \citenamefont {Brandenburg}, \citenamefont {Rogachevskii},\ and\
  \citenamefont {Kleeorin}}]{SBR19}%
  \BibitemOpen
  \bibfield  {author} {\bibinfo {author} {\bibfnamefont {J.}~\bibnamefont
  {Schober}}, \bibinfo {author} {\bibfnamefont {A.}~\bibnamefont
  {Brandenburg}}, \bibinfo {author} {\bibfnamefont {I.}~\bibnamefont
  {Rogachevskii}},\ and\ \bibinfo {author} {\bibfnamefont {N.}~\bibnamefont
  {Kleeorin}},\ }\bibfield  {title} {\bibinfo {title} {Energetics of turbulence
  generated by chiral mhd dynamos},\ }\href@noop {} {\bibfield  {journal}
  {\bibinfo  {journal} {Geophys. Astrophys. Fluid Dyn.}\ }\textbf {\bibinfo
  {volume} {113}},\ \bibinfo {pages} {107} (\bibinfo {year}
  {2019})}\BibitemShut {NoStop}%
\bibitem [{\citenamefont {Schober}\ \emph {et~al.}(2020)\citenamefont
  {Schober}, \citenamefont {Brandenburg},\ and\ \citenamefont
  {Rogachevskii}}]{SBR20}%
  \BibitemOpen
  \bibfield  {author} {\bibinfo {author} {\bibfnamefont {J.}~\bibnamefont
  {Schober}}, \bibinfo {author} {\bibfnamefont {A.}~\bibnamefont
  {Brandenburg}},\ and\ \bibinfo {author} {\bibfnamefont {I.}~\bibnamefont
  {Rogachevskii}},\ }\bibfield  {title} {\bibinfo {title} {Chiral fermion
  asymmetry in high-energy plasma simulations},\ }\href@noop {} {\bibfield
  {journal} {\bibinfo  {journal} {Geophys. Astrophys. Fluid Dyn.}\ }\textbf
  {\bibinfo {volume} {114}},\ \bibinfo {pages} {106} (\bibinfo {year}
  {2020})}\BibitemShut {NoStop}%
\bibitem [{\citenamefont {{Joyce}}\ and\ \citenamefont
  {{Shaposhnikov}}(1997)}]{JS97}%
  \BibitemOpen
  \bibfield  {author} {\bibinfo {author} {\bibfnamefont {M.}~\bibnamefont
  {{Joyce}}}\ and\ \bibinfo {author} {\bibfnamefont {M.}~\bibnamefont
  {{Shaposhnikov}}},\ }\bibfield  {title} {\bibinfo {title} {{Primordial
  Magnetic Fields, Right Electrons, and the Abelian Anomaly}},\ }\href@noop {}
  {\bibfield  {journal} {\bibinfo  {journal} {Phys.~Rev.~Lett.}\ }\textbf
  {\bibinfo {volume} {79}},\ \bibinfo {pages} {1193} (\bibinfo {year}
  {1997})}\BibitemShut {NoStop}%
\bibitem [{\citenamefont {{Boyarsky}}\ \emph {et~al.}(2012)\citenamefont
  {{Boyarsky}}, \citenamefont {{Fr{\"o}hlich}},\ and\ \citenamefont
  {{Ruchayskiy}}}]{BFR12}%
  \BibitemOpen
  \bibfield  {author} {\bibinfo {author} {\bibfnamefont {A.}~\bibnamefont
  {{Boyarsky}}}, \bibinfo {author} {\bibfnamefont {J.}~\bibnamefont
  {{Fr{\"o}hlich}}},\ and\ \bibinfo {author} {\bibfnamefont {O.}~\bibnamefont
  {{Ruchayskiy}}},\ }\bibfield  {title} {\bibinfo {title} {{Self-consistent
  evolution of magnetic fields and chiral asymmetry in the early universe}},\
  }\href@noop {} {\bibfield  {journal} {\bibinfo  {journal} {Phys. Rev. Lett.}\
  }\textbf {\bibinfo {volume} {108}} (\bibinfo {year} {2012})}\BibitemShut
  {NoStop}%
\bibitem [{\citenamefont {Boyarsky}\ \emph {et~al.}(2012)\citenamefont
  {Boyarsky}, \citenamefont {Ruchayskiy},\ and\ \citenamefont
  {Shaposhnikov}}]{BRS12}%
  \BibitemOpen
  \bibfield  {author} {\bibinfo {author} {\bibfnamefont {A.}~\bibnamefont
  {Boyarsky}}, \bibinfo {author} {\bibfnamefont {O.}~\bibnamefont
  {Ruchayskiy}},\ and\ \bibinfo {author} {\bibfnamefont {M.}~\bibnamefont
  {Shaposhnikov}},\ }\bibfield  {title} {\bibinfo {title} {Long-range magnetic
  fields in the ground state of the standard model plasma},\ }\href@noop {}
  {\bibfield  {journal} {\bibinfo  {journal} {Phys. Rev. Lett.}\ }\textbf
  {\bibinfo {volume} {109}},\ \bibinfo {pages} {111602} (\bibinfo {year}
  {2012})}\BibitemShut {NoStop}%
\bibitem [{\citenamefont {{Dvornikov}}\ and\ \citenamefont
  {{Semikoz}}(2017)}]{DvornikovSemikoz2017}%
  \BibitemOpen
  \bibfield  {author} {\bibinfo {author} {\bibfnamefont {M.}~\bibnamefont
  {{Dvornikov}}}\ and\ \bibinfo {author} {\bibfnamefont {V.~B.}\ \bibnamefont
  {{Semikoz}}},\ }\bibfield  {title} {\bibinfo {title} {{Influence of the
  turbulent motion on the chiral magnetic effect in the early universe}},\
  }\href {https://doi.org/10.1103/PhysRevD.95.043538} {\bibfield  {journal}
  {\bibinfo  {journal} {\prd}\ }\textbf {\bibinfo {volume} {95}},\ \bibinfo
  {eid} {043538} (\bibinfo {year} {2017})}\BibitemShut {NoStop}%
\bibitem [{\citenamefont {{Brandenburg}}\ \emph {et~al.}(2017)\citenamefont
  {{Brandenburg}}, \citenamefont {{Schober}}, \citenamefont {{Rogachevskii}},
  \citenamefont {{Kahniashvili}}, \citenamefont {{Boyarsky}}, \citenamefont
  {{Fr\"ohlich}}, \citenamefont {{Ruchayskiy}},\ and\ \citenamefont
  {{Kleeorin}}}]{BSRKBFRK17}%
  \BibitemOpen
  \bibfield  {author} {\bibinfo {author} {\bibfnamefont {A.}~\bibnamefont
  {{Brandenburg}}}, \bibinfo {author} {\bibfnamefont {J.}~\bibnamefont
  {{Schober}}}, \bibinfo {author} {\bibfnamefont {I.}~\bibnamefont
  {{Rogachevskii}}}, \bibinfo {author} {\bibfnamefont {T.}~\bibnamefont
  {{Kahniashvili}}}, \bibinfo {author} {\bibfnamefont {A.}~\bibnamefont
  {{Boyarsky}}}, \bibinfo {author} {\bibfnamefont {J.}~\bibnamefont
  {{Fr\"ohlich}}}, \bibinfo {author} {\bibfnamefont {O.}~\bibnamefont
  {{Ruchayskiy}}},\ and\ \bibinfo {author} {\bibfnamefont {N.}~\bibnamefont
  {{Kleeorin}}},\ }\bibfield  {title} {\bibinfo {title} {{The turbulent
  chiral-magnetic cascade in the early universe}},\ }\href@noop {} {\bibfield
  {journal} {\bibinfo  {journal} {Astrophys. J. Lett.}\ }\textbf {\bibinfo
  {volume} {845}},\ \bibinfo {eid} {L21} (\bibinfo {year} {2017})},\ \Eprint
  {https://arxiv.org/abs/1707.03385} {arXiv:1707.03385} \BibitemShut {NoStop}%
\bibitem [{\citenamefont {{Dvornikov}}\ and\ \citenamefont
  {{Semikoz}}(2015)}]{Dvornikov:2015lea}%
  \BibitemOpen
  \bibfield  {author} {\bibinfo {author} {\bibfnamefont {M.}~\bibnamefont
  {{Dvornikov}}}\ and\ \bibinfo {author} {\bibfnamefont {V.~B.}\ \bibnamefont
  {{Semikoz}}},\ }\bibfield  {title} {\bibinfo {title} {{Generation of the
  magnetic helicity in a neutron star driven by the electroweak
  electron-nucleon interaction}},\ }\href
  {https://doi.org/10.1088/1475-7516/2015/05/032} {\bibfield  {journal}
  {\bibinfo  {journal} {J. Cosmol. Astropart. Phys}\ }\textbf {\bibinfo
  {volume} {05}},\ \bibinfo {pages} {032} (\bibinfo {year} {2015})},\ \Eprint
  {https://arxiv.org/abs/1503.04162} {arXiv:1503.04162 [astro-ph.HE]}
  \BibitemShut {NoStop}%
\bibitem [{\citenamefont {Yamamoto}(2016)}]{YA16}%
  \BibitemOpen
  \bibfield  {author} {\bibinfo {author} {\bibfnamefont {N.}~\bibnamefont
  {Yamamoto}},\ }\bibfield  {title} {\bibinfo {title} {Scaling laws in chiral
  hydrodynamic turbulence},\ }\href@noop {} {\bibfield  {journal} {\bibinfo
  {journal} {Phys. Rev. D}\ }\textbf {\bibinfo {volume} {93}},\ \bibinfo
  {pages} {125016} (\bibinfo {year} {2016})}\BibitemShut {NoStop}%
\bibitem [{\citenamefont {Sigl}\ and\ \citenamefont
  {Leite}(2016)}]{Sigl:2015xva}%
  \BibitemOpen
  \bibfield  {author} {\bibinfo {author} {\bibfnamefont {G.}~\bibnamefont
  {Sigl}}\ and\ \bibinfo {author} {\bibfnamefont {N.}~\bibnamefont {Leite}},\
  }\bibfield  {title} {\bibinfo {title} {{Chiral magnetic effect in
  protoneutron stars and magnetic field spectral evolution}},\ }\href
  {https://doi.org/10.1088/1475-7516/2016/01/025} {\bibfield  {journal}
  {\bibinfo  {journal} {J. Cosmol. Astropart. Phys}\ }\textbf {\bibinfo
  {volume} {01}},\ \bibinfo {pages} {025} (\bibinfo {year} {2016})},\ \Eprint
  {https://arxiv.org/abs/1507.04983} {arXiv:1507.04983 [astro-ph.HE]}
  \BibitemShut {NoStop}%
\bibitem [{\citenamefont {Akamatsu}\ and\ \citenamefont
  {Yamamoto}(2013)}]{AY13}%
  \BibitemOpen
  \bibfield  {author} {\bibinfo {author} {\bibfnamefont {Y.}~\bibnamefont
  {Akamatsu}}\ and\ \bibinfo {author} {\bibfnamefont {N.}~\bibnamefont
  {Yamamoto}},\ }\bibfield  {title} {\bibinfo {title} {Chiral plasma
  instabilities},\ }\href@noop {} {\bibfield  {journal} {\bibinfo  {journal}
  {Phys. Rev. Lett.}\ }\textbf {\bibinfo {volume} {111}},\ \bibinfo {pages}
  {052002} (\bibinfo {year} {2013})}\BibitemShut {NoStop}%
\bibitem [{\citenamefont {Hirono}\ \emph {et~al.}(2015)\citenamefont {Hirono},
  \citenamefont {Kharzeev},\ and\ \citenamefont {Yin}}]{HKY15}%
  \BibitemOpen
  \bibfield  {author} {\bibinfo {author} {\bibfnamefont {Y.}~\bibnamefont
  {Hirono}}, \bibinfo {author} {\bibfnamefont {D.~E.}\ \bibnamefont
  {Kharzeev}},\ and\ \bibinfo {author} {\bibfnamefont {Y.}~\bibnamefont
  {Yin}},\ }\bibfield  {title} {\bibinfo {title} {Self-similar inverse cascade
  of magnetic helicity driven by the chiral anomaly},\ }\href@noop {}
  {\bibfield  {journal} {\bibinfo  {journal} {Phys. Rev. D}\ }\textbf {\bibinfo
  {volume} {92}},\ \bibinfo {pages} {125031} (\bibinfo {year}
  {2015})}\BibitemShut {NoStop}%
\bibitem [{\citenamefont {Taghavi}\ and\ \citenamefont
  {Wiedemann}(2015)}]{TW15}%
  \BibitemOpen
  \bibfield  {author} {\bibinfo {author} {\bibfnamefont {S.~F.}\ \bibnamefont
  {Taghavi}}\ and\ \bibinfo {author} {\bibfnamefont {U.~A.}\ \bibnamefont
  {Wiedemann}},\ }\bibfield  {title} {\bibinfo {title} {Chiral magnetic wave in
  an expanding qcd fluid},\ }\href@noop {} {\bibfield  {journal} {\bibinfo
  {journal} {Phys. Rev. C}\ }\textbf {\bibinfo {volume} {91}},\ \bibinfo
  {pages} {024902} (\bibinfo {year} {2015})}\BibitemShut {NoStop}%
\bibitem [{\citenamefont {Miransky}\ and\ \citenamefont
  {Shovkovy}(2015)}]{MS15}%
  \BibitemOpen
  \bibfield  {author} {\bibinfo {author} {\bibfnamefont {V.~A.}\ \bibnamefont
  {Miransky}}\ and\ \bibinfo {author} {\bibfnamefont {I.~A.}\ \bibnamefont
  {Shovkovy}},\ }\bibfield  {title} {\bibinfo {title} {Quantum field theory in
  a magnetic field: From quantum chromodynamics to graphene and dirac
  semimetals},\ }\href@noop {} {\bibfield  {journal} {\bibinfo  {journal}
  {Phys. Rep.}\ }\textbf {\bibinfo {volume} {576}},\ \bibinfo {pages} {1}
  (\bibinfo {year} {2015})}\BibitemShut {NoStop}%
\bibitem [{\citenamefont {Schober}\ \emph
  {et~al.}(2022{\natexlab{a}})\citenamefont {Schober}, \citenamefont
  {Rogachevskii},\ and\ \citenamefont {Brandenburg}}]{SchoberEtAl2022a}%
  \BibitemOpen
  \bibfield  {author} {\bibinfo {author} {\bibfnamefont {J.}~\bibnamefont
  {Schober}}, \bibinfo {author} {\bibfnamefont {I.}~\bibnamefont
  {Rogachevskii}},\ and\ \bibinfo {author} {\bibfnamefont {A.}~\bibnamefont
  {Brandenburg}},\ }\bibfield  {title} {\bibinfo {title} {Production of a
  chiral magnetic anomaly with emerging turbulence and mean-field dynamo
  action},\ }\href {https://doi.org/10.1103/PhysRevLett.128.065002} {\bibfield
  {journal} {\bibinfo  {journal} {Phys. Rev. Lett.}\ }\textbf {\bibinfo
  {volume} {128}},\ \bibinfo {pages} {065002} (\bibinfo {year}
  {2022}{\natexlab{a}})}\BibitemShut {NoStop}%
\bibitem [{\citenamefont {Schober}\ \emph
  {et~al.}(2022{\natexlab{b}})\citenamefont {Schober}, \citenamefont
  {Rogachevskii},\ and\ \citenamefont {Brandenburg}}]{SchoberEtAl2022b}%
  \BibitemOpen
  \bibfield  {author} {\bibinfo {author} {\bibfnamefont {J.}~\bibnamefont
  {Schober}}, \bibinfo {author} {\bibfnamefont {I.}~\bibnamefont
  {Rogachevskii}},\ and\ \bibinfo {author} {\bibfnamefont {A.}~\bibnamefont
  {Brandenburg}},\ }\bibfield  {title} {\bibinfo {title} {Dynamo instabilities
  in plasmas with inhomogeneous chiral chemical potential},\ }\href
  {https://doi.org/10.1103/PhysRevD.105.043507} {\bibfield  {journal} {\bibinfo
   {journal} {Phys. Rev. D}\ }\textbf {\bibinfo {volume} {105}},\ \bibinfo
  {pages} {043507} (\bibinfo {year} {2022}{\natexlab{b}})}\BibitemShut
  {NoStop}%
\bibitem [{\citenamefont {{Kharzeev}}\ and\ \citenamefont
  {{Yee}}(2011)}]{KY11}%
  \BibitemOpen
  \bibfield  {author} {\bibinfo {author} {\bibfnamefont {D.~E.}\ \bibnamefont
  {{Kharzeev}}}\ and\ \bibinfo {author} {\bibfnamefont {H.-U.}\ \bibnamefont
  {{Yee}}},\ }\bibfield  {title} {\bibinfo {title} {{Chiral magnetic wave}},\
  }\href {https://doi.org/10.1103/PhysRevD.83.085007} {\bibfield  {journal}
  {\bibinfo  {journal} {\prd}\ }\textbf {\bibinfo {volume} {83}},\ \bibinfo
  {eid} {085007} (\bibinfo {year} {2011})},\ \Eprint
  {https://arxiv.org/abs/1012.6026} {arXiv:1012.6026 [hep-th]} \BibitemShut
  {NoStop}%
\bibitem [{\citenamefont {{Pencil Code Collaboration}}\ \emph
  {et~al.}(2021)\citenamefont {{Pencil Code Collaboration}}, \citenamefont
  {{Brandenburg}}, \citenamefont {{Johansen}}, \citenamefont {{Bourdin}},
  \citenamefont {{Dobler}}, \citenamefont {{Lyra}}, \citenamefont
  {{Rheinhardt}}, \citenamefont {{Bingert}}, \citenamefont {{Haugen}},
  \citenamefont {{Mee}}, \citenamefont {{Gent}}, \citenamefont {{Babkovskaia}},
  \citenamefont {{Yang}}, \citenamefont {{Heinemann}}, \citenamefont
  {{Dintrans}}, \citenamefont {{Mitra}}, \citenamefont {{Candelaresi}},
  \citenamefont {{Warnecke}}, \citenamefont {{K{\"a}pyl{\"a}}}, \citenamefont
  {{Schreiber}}, \citenamefont {{Chatterjee}}, \citenamefont
  {{K{\"a}pyl{\"a}}}, \citenamefont {{Li}}, \citenamefont {{Kr{\"u}ger}},
  \citenamefont {{Aarnes}}, \citenamefont {{Sarson}}, \citenamefont {{Oishi}},
  \citenamefont {{Schober}}, \citenamefont {{Plasson}}, \citenamefont
  {{Sandin}}, \citenamefont {{Karchniwy}}, \citenamefont {{Rodrigues}},
  \citenamefont {{Hubbard}}, \citenamefont {{Guerrero}}, \citenamefont
  {{Snodin}}, \citenamefont {{Losada}}, \citenamefont {{Pekkil{\"a}}},\ and\
  \citenamefont {{Qian}}}]{PencilCode2021}%
  \BibitemOpen
  \bibfield  {author} {\bibinfo {author} {\bibnamefont {{Pencil Code
  Collaboration}}}, \bibinfo {author} {\bibfnamefont {A.}~\bibnamefont
  {{Brandenburg}}}, \bibinfo {author} {\bibfnamefont {A.}~\bibnamefont
  {{Johansen}}}, \bibinfo {author} {\bibfnamefont {P.}~\bibnamefont
  {{Bourdin}}}, \bibinfo {author} {\bibfnamefont {W.}~\bibnamefont {{Dobler}}},
  \bibinfo {author} {\bibfnamefont {W.}~\bibnamefont {{Lyra}}}, \bibinfo
  {author} {\bibfnamefont {M.}~\bibnamefont {{Rheinhardt}}}, \bibinfo {author}
  {\bibfnamefont {S.}~\bibnamefont {{Bingert}}}, \bibinfo {author}
  {\bibfnamefont {N.}~\bibnamefont {{Haugen}}}, \bibinfo {author}
  {\bibfnamefont {A.}~\bibnamefont {{Mee}}}, \bibinfo {author} {\bibfnamefont
  {F.}~\bibnamefont {{Gent}}}, \bibinfo {author} {\bibfnamefont
  {N.}~\bibnamefont {{Babkovskaia}}}, \bibinfo {author} {\bibfnamefont {C.-C.}\
  \bibnamefont {{Yang}}}, \bibinfo {author} {\bibfnamefont {T.}~\bibnamefont
  {{Heinemann}}}, \bibinfo {author} {\bibfnamefont {B.}~\bibnamefont
  {{Dintrans}}}, \bibinfo {author} {\bibfnamefont {D.}~\bibnamefont {{Mitra}}},
  \bibinfo {author} {\bibfnamefont {S.}~\bibnamefont {{Candelaresi}}}, \bibinfo
  {author} {\bibfnamefont {J.}~\bibnamefont {{Warnecke}}}, \bibinfo {author}
  {\bibfnamefont {P.}~\bibnamefont {{K{\"a}pyl{\"a}}}}, \bibinfo {author}
  {\bibfnamefont {A.}~\bibnamefont {{Schreiber}}}, \bibinfo {author}
  {\bibfnamefont {P.}~\bibnamefont {{Chatterjee}}}, \bibinfo {author}
  {\bibfnamefont {M.}~\bibnamefont {{K{\"a}pyl{\"a}}}}, \bibinfo {author}
  {\bibfnamefont {X.-Y.}\ \bibnamefont {{Li}}}, \bibinfo {author}
  {\bibfnamefont {J.}~\bibnamefont {{Kr{\"u}ger}}}, \bibinfo {author}
  {\bibfnamefont {J.}~\bibnamefont {{Aarnes}}}, \bibinfo {author}
  {\bibfnamefont {G.}~\bibnamefont {{Sarson}}}, \bibinfo {author}
  {\bibfnamefont {J.}~\bibnamefont {{Oishi}}}, \bibinfo {author} {\bibfnamefont
  {J.}~\bibnamefont {{Schober}}}, \bibinfo {author} {\bibfnamefont
  {R.}~\bibnamefont {{Plasson}}}, \bibinfo {author} {\bibfnamefont
  {C.}~\bibnamefont {{Sandin}}}, \bibinfo {author} {\bibfnamefont
  {E.}~\bibnamefont {{Karchniwy}}}, \bibinfo {author} {\bibfnamefont
  {L.}~\bibnamefont {{Rodrigues}}}, \bibinfo {author} {\bibfnamefont
  {A.}~\bibnamefont {{Hubbard}}}, \bibinfo {author} {\bibfnamefont
  {G.}~\bibnamefont {{Guerrero}}}, \bibinfo {author} {\bibfnamefont
  {A.}~\bibnamefont {{Snodin}}}, \bibinfo {author} {\bibfnamefont
  {I.}~\bibnamefont {{Losada}}}, \bibinfo {author} {\bibfnamefont
  {J.}~\bibnamefont {{Pekkil{\"a}}}},\ and\ \bibinfo {author} {\bibfnamefont
  {C.}~\bibnamefont {{Qian}}},\ }\bibfield  {title} {\bibinfo {title} {{The
  Pencil Code, a modular MPI code for partial differential equations and
  particles: multipurpose and multiuser-maintained}},\ }\href
  {https://doi.org/10.21105/joss.02807} {\bibfield  {journal} {\bibinfo
  {journal} {J. Open Source Software}\ }\textbf {\bibinfo {volume} {6}},\
  \bibinfo {eid} {2807} (\bibinfo {year} {2021})},\ \Eprint
  {https://arxiv.org/abs/2009.08231} {arXiv:2009.08231 [astro-ph.IM]}
  \BibitemShut {NoStop}%
\bibitem [{\citenamefont {{Williamson}}(1980)}]{Wil80}%
  \BibitemOpen
  \bibfield  {author} {\bibinfo {author} {\bibfnamefont {J.~H.}\ \bibnamefont
  {{Williamson}}},\ }\bibfield  {title} {\bibinfo {title} {{Low-storage
  Runge-Kutta schemes}},\ }\href@noop {} {\bibfield  {journal} {\bibinfo
  {journal} {J. Comp. Phys.}\ }\textbf {\bibinfo {volume} {35}},\ \bibinfo
  {pages} {48} (\bibinfo {year} {1980})}\BibitemShut {NoStop}%
\bibitem [{\citenamefont {{Brandenburg}}\ and\ \citenamefont
  {{Dobler}}(2002)}]{BD02}%
  \BibitemOpen
  \bibfield  {author} {\bibinfo {author} {\bibfnamefont {A.}~\bibnamefont
  {{Brandenburg}}}\ and\ \bibinfo {author} {\bibfnamefont {W.}~\bibnamefont
  {{Dobler}}},\ }\bibfield  {title} {\bibinfo {title} {{Hydromagnetic
  turbulence in computer simulations}},\ }\href
  {https://doi.org/10.1016/S0010-4655(02)00334-X} {\bibfield  {journal}
  {\bibinfo  {journal} {Comp. Phys. Comm.}\ }\textbf {\bibinfo {volume}
  {147}},\ \bibinfo {pages} {471} (\bibinfo {year} {2002})},\ \Eprint
  {https://arxiv.org/abs/astro-ph/0111569} {astro-ph/0111569} \BibitemShut
  {NoStop}%
\bibitem [{\citenamefont {Brandenburg}(2003)}]{Bra03}%
  \BibitemOpen
  \bibfield  {author} {\bibinfo {author} {\bibfnamefont {A.}~\bibnamefont
  {Brandenburg}},\ }\bibfield  {title} {\bibinfo {title} {Computational aspects
  of astrophysical mhd and turbulence},\ }in\ \href
  {https://doi.org/10.1201/9780203493137.ch9} {\emph {\bibinfo {booktitle}
  {Advances in Nonlinear Dynamics}}},\ \bibinfo {editor} {edited by\ \bibinfo
  {editor} {\bibfnamefont {A.}~\bibnamefont {{Ferriz-Mas}}}\ and\ \bibinfo
  {editor} {\bibfnamefont {M.}~\bibnamefont {{N{\'u}{\~n}ez}}}}\ (\bibinfo
  {publisher} {CRC Press},\ \bibinfo {year} {2003})\ pp.\ \bibinfo {pages}
  {269--344}\BibitemShut {NoStop}%
\bibitem [{\citenamefont {{Schober}}\ \emph {et~al.}(2023)\citenamefont
  {{Schober}}, \citenamefont {{Rogachevskii}},\ and\ \citenamefont
  {{Brandenburg}}}]{SchoberEtAl2023b}%
  \BibitemOpen
  \bibfield  {author} {\bibinfo {author} {\bibfnamefont {J.}~\bibnamefont
  {{Schober}}}, \bibinfo {author} {\bibfnamefont {I.}~\bibnamefont
  {{Rogachevskii}}},\ and\ \bibinfo {author} {\bibfnamefont {A.}~\bibnamefont
  {{Brandenburg}}},\ }\bibfield  {title} {\bibinfo {title} {{Efficiency of
  dynamos from autonomous generation of chiral asymmetry}},\ }\href@noop {}
  {\bibfield  {journal} {\bibinfo  {journal} {unpublished}\ } (\bibinfo {year}
  {2023})}\BibitemShut {NoStop}%
\bibitem [{\citenamefont {{Subramanian}}\ and\ \citenamefont
  {{Brandenburg}}(2014)}]{SubramanianBrandenburg2014}%
  \BibitemOpen
  \bibfield  {author} {\bibinfo {author} {\bibfnamefont {K.}~\bibnamefont
  {{Subramanian}}}\ and\ \bibinfo {author} {\bibfnamefont {A.}~\bibnamefont
  {{Brandenburg}}},\ }\bibfield  {title} {\bibinfo {title} {{Traces of
  large-scale dynamo action in the kinematic stage}},\ }\href
  {https://doi.org/10.1093/mnras/stu1954} {\bibfield  {journal} {\bibinfo
  {journal} {\mnras}\ }\textbf {\bibinfo {volume} {445}},\ \bibinfo {pages}
  {2930} (\bibinfo {year} {2014})},\ \Eprint {https://arxiv.org/abs/1408.4416}
  {arXiv:1408.4416 [astro-ph.GA]} \BibitemShut {NoStop}%
\bibitem [{\citenamefont {{Hosking}}\ and\ \citenamefont
  {{Schekochihin}}(2021)}]{HS21}%
  \BibitemOpen
  \bibfield  {author} {\bibinfo {author} {\bibfnamefont {D.~N.}\ \bibnamefont
  {{Hosking}}}\ and\ \bibinfo {author} {\bibfnamefont {A.~A.}\ \bibnamefont
  {{Schekochihin}}},\ }\bibfield  {title} {\bibinfo {title}
  {{Reconnection-Controlled Decay of Magnetohydrodynamic Turbulence and the
  Role of Invariants}},\ }\href {https://doi.org/10.1103/PhysRevX.11.041005}
  {\bibfield  {journal} {\bibinfo  {journal} {Phys. Rev. X}\ }\textbf {\bibinfo
  {volume} {11}},\ \bibinfo {eid} {041005} (\bibinfo {year} {2021})},\ \Eprint
  {https://arxiv.org/abs/2012.01393} {arXiv:2012.01393 [physics.flu-dyn]}
  \BibitemShut {NoStop}%
\bibitem [{\citenamefont {Brandenburg}\ \emph {et~al.}(2023)\citenamefont
  {Brandenburg}, \citenamefont {Kamada},\ and\ \citenamefont
  {Schober}}]{BKS2023}%
  \BibitemOpen
  \bibfield  {author} {\bibinfo {author} {\bibfnamefont {A.}~\bibnamefont
  {Brandenburg}}, \bibinfo {author} {\bibfnamefont {K.}~\bibnamefont
  {Kamada}},\ and\ \bibinfo {author} {\bibfnamefont {J.}~\bibnamefont
  {Schober}},\ }\bibfield  {title} {\bibinfo {title} {Decay law of magnetic
  turbulence with helicity balanced by chiral fermions},\ }\href
  {https://doi.org/10.1103/PhysRevResearch.5.L022028} {\bibfield  {journal}
  {\bibinfo  {journal} {Phys. Rev. Res.}\ }\textbf {\bibinfo {volume} {5}},\
  \bibinfo {pages} {L022028} (\bibinfo {year} {2023})}\BibitemShut {NoStop}%
\bibitem [{\citenamefont {{Sigl}}\ and\ \citenamefont
  {{Leite}}(2016)}]{SiglLeite2016}%
  \BibitemOpen
  \bibfield  {author} {\bibinfo {author} {\bibfnamefont {G.}~\bibnamefont
  {{Sigl}}}\ and\ \bibinfo {author} {\bibfnamefont {N.}~\bibnamefont
  {{Leite}}},\ }\bibfield  {title} {\bibinfo {title} {{Chiral magnetic effect
  in protoneutron stars and magnetic field spectral evolution}},\ }\href
  {https://doi.org/10.1088/1475-7516/2016/01/025} {\bibfield  {journal}
  {\bibinfo  {journal} {J. Cosmol. Astropart. Physics}\ }\textbf {\bibinfo
  {volume} {1}},\ \bibinfo {eid} {025} (\bibinfo {year} {2016})}\BibitemShut
  {NoStop}%
\end{thebibliography}
\end{document}